\documentclass[]{aastex631}

\shorttitle{Revisiting the XRB MAXI J1631--479}
\shortauthors{Rout et al.}

\begin{document}

\title{Revisiting the galactic X-ray binary MAXI J1631--479:
Implications for high inclination and a massive black hole}

\correspondingauthor{Sandeep K. Rout}
\email{skrout@prl.res.in, skrout92@gmail.com}

\author[ 0000-0001-7590-5099 ]{Sandeep K. Rout}
\affiliation{Physical Research Laboratory, Ahmedabad, Gujarat 380009, India}

\author[ 0000-0002-2050-0913 ]{Santosh Vadawale}
\affiliation{Physical Research Laboratory, Ahmedabad, Gujarat 380009, India}

\author[ 0000-0003-3828-2448 ]{Javier Gar\'cia}
\affiliation{Cahill Center for Astronomy and Astrophysics, California Institute of Technology, Pasadena, CA 91125, USA}

\author[ 0000-0002-8908-759X ]{Riley Connors}
\affiliation{Cahill Center for Astronomy and Astrophysics, California Institute of Technology, Pasadena, CA 91125, USA}

\begin{abstract}

X-ray spectroscopy of galactic black hole binaries serve as a powerful tool to gain an overall understanding of the system. Not only can the properties of the accretion disk be studied in detail, the fundamental properties of the black hole can also be inferred. The pursuit of these objectives also leads to an indirect validation of general relativity in strong field limit. In this work we carry out a comprehensive spectral analysis of the galactic X-ray binary MAXI J1631--479 using data from \textit{NICER} and \textit{NuSTAR} observatories. We trace the evolution of the accretion disk properties such as density, ionization, Fe abundance, etc as the source transitions from a disk dominated soft state to a power law dominated hard intermediate state. We provide strong constrains on the spin of the black hole and the inclination of the inner disk. We also use the soft state \textit{NICER} observations to constrain the black hole mass using distance estimates from optical observations.

\end{abstract}


\section{Introduction} \label{sec:intro}

X-rays emitted from black-hole binaries contain a plethora of information about the system which can be extracted using various techniques. In order to understand the various system properties using spectroscopy it is essential to accurately predict the emission mechanisms. The most common source of soft X-ray emission is blackbody radiation from a geometrically thin and optically thick accretion disk \citep{shakura73, novikov73}. This thermal emission undergoes inverse Compton scattering from a hot electron cloud (Corona) located in the vicinity of the black hole \citep[][ and references therein]{done07}. The Compton spectrum manifests as a powerlaw with a cut-off at lower energy and an exponential roll over at higher energies. While the low energy cut-off is controlled by the seed photon energy, the high energy roll over is driven by the maximum energy of the electrons in the Corona. When a fraction of the Compton upscattered hard X-rays move down and interact with the thin accretion disk, they get absorbed and scattered resulting in a reflection component. This component has two prominent emission features - a smeared fluorescent Fe K$\alpha$ line and a broad Compton hump between 10 - 30 keV \citep{fabian89}. 

Once the emission mechanisms are pinned down several useful information can be drawn by accurate modeling of the spectrum. An often undertaken exercise is the estimation of black hole spin and disk inclination using relativistic reflection spectroscopy \citep{fabian89,miller07}. The red and blue shift of the fluorescent Fe line provides constraints on the  spin and inclination \citep{fabian00}. Other variable parameters like density, Fe abundance, and ionization of the disk can also be estimated from reflection modeling \citep{garcia16, tomsick18}. Complexities in the data often pave ways for new interpretations which are not mainstream such as emission from the plunging region \citep{fabian20} and reflection by returning radiation \citep{connors20}. 

The mass of the black hole is another important parameter which along with the spin can fully characterize all its observable properties. The conventional method for measuring black hole mass hosted in X-ray binaries involves radial velocity measurement of emission lines obtained from the companion star. Apart from this method, the mass can also be constrained using X-ray emission from inner most regions of the accretion disk during the high/soft state. The soft X-ray flux ($\sim 0.1 - 3$ keV) is a function of black hole mass along with distance, spin, and inclination \citep{zhang97}. A robust estimate of the spin and inclination (from reflection spectroscopy) can provide a relation between mass and distance \citep{parker16}. A limit on the distance can consequently provide a limit on mass. We use this technique to constrain the black hole mass for a newly discovered binary MAXI J1631--479 (hereafter, J1631).

J1631 was detected as a bright hard X-ray transient with \textit{MAXI}/GSC on 2018 December 21 at 04:33 UTC. The outburst was marked by a fast rise in luminosity till 2019 January 07 followed by an exponential decay, a pattern which is typical of black hole binaries. \citet{fiocchi20} reported results on the \textit{INTEGRAL}/IBIS spectrum suggesting possible emission from a hybrid plasma. \citet{xu20} studied the variation of reflection features using \textit{NuSTAR} as the source transitioned from a disk-dominated state to a powerlaw-dominated state. \citet{rout21a} studied the spectral and timing evolution of the source using \textit{NICER}.

\begin{figure}
\centering
    \includegraphics[scale=0.7]{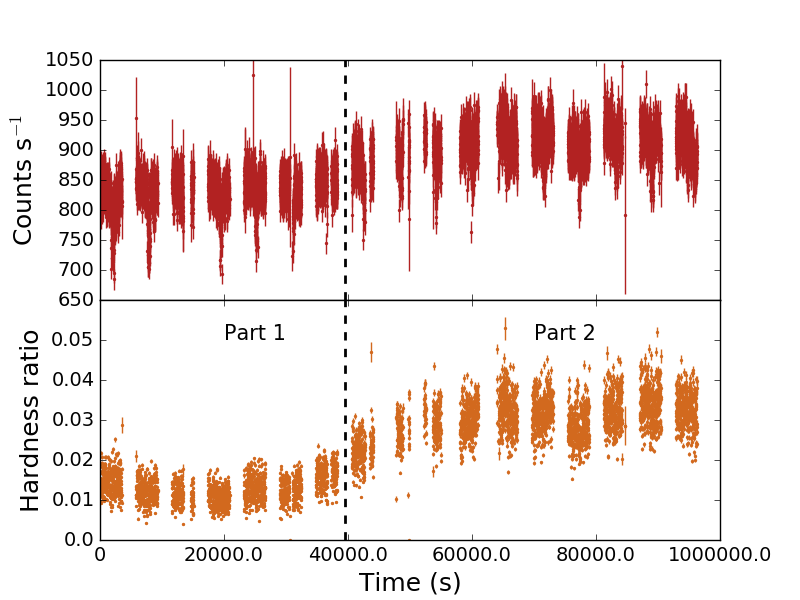}
    \caption{\textit{NuSTAR} light curve and hardness ratio for the January 17 observation. The hardness ratio is defined as the ratio of count rate in 10 - 78 keV to that in 3 - 10 keV. The two parts represent two different levels of count rate and hardness and thus, the spectral analysis was done separately for each part. Only data from FPMA is shown here.}
    \label{fig:hss_lc}
\end{figure}

\section{Observation \& Data reduction}

Three epochs of archival \textit{NuSTAR} \citep{harrison13} observations made on 2019 January 17, 27, and 30 (ObsIDs: 90501301001, 80401316002, and 80401316004) were analyzed in this work. These data were reduced using the \textit{NuSTAR} Data Analysis Software-v2.1.1 and the calibration database v20220118 using the standard procedure. The source spectra were extracted from circular regions of 200$''$ around the point spread function and the background spectra were extracted from circular source-free regions of appropriate sizes. The January 17 observation was divided into two parts as there is an increase in count rate and hardness in the second half of the observation (see Figure 1 of \citet{xu20}). The spectrum from the first part of the January 17 observation was restricted up to 40 keV as the background dominated beyond it. All spectra were grouped to have a minimum of 25 counts per energy bin to facilitate $\chi^2$ statistics. We also repeated the spectral analysis by binning the spectra to have a minimum signal to noise ratio of 3.0 and also optimally following \citet{kaastra16}. It was verified that the results of the analysis do not vary with the binning scheme opted. While the January 17 observation was during the high soft state (HSS), the January 27 \& 30 observations were during the hard intermediate state (HIMS) \citep{rout21a} \citep[see][for a definition of various states]{belloni10}. In order to study the spectral evolution, the spectra during these states were fitted simultaneously in two groups. Henceforth, the two groups will be identified as the HSS and the HIMS spectra.

One of the objectives of this paper is to constrain the mass of the black hole in J1631 by modeling the thermal disk emission during the HSS \citep{parker16}. J1631 remained in the HSS till 2019 January 23 \citep{rout21a}. Therefore, data from the first nine \textit{NICER} \citep{gendreau16} observations (from ObsID 1200500101 to 1200500109) were reduced using the recent versions of the pipeline (2021-08-31\_V008c), calibration database (v20210707), and background model (\texttt{nibackgen3C50-v7b}). Two other observations during the HIMS (ObsID 1200500114 and 1200500126) were also analyzed (see Sections \ref{hims_ana} and \ref{mass_ana} for details). The spectra were grouped to have a minimum of 25 counts per bin and a systematic error of 1\% was added while fitting. Other steps of analysis were similar to those mentioned in \citet{rout21a}

\section{Analysis \& Results}

\begin{figure}
\centering
    \includegraphics[scale=0.6]{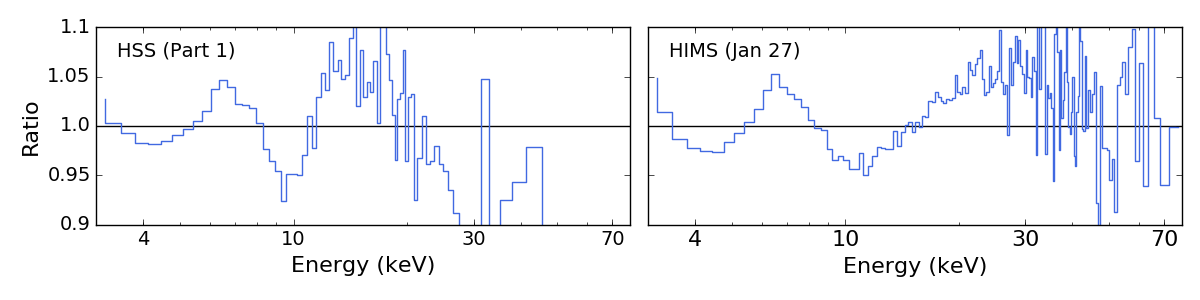}
    \caption{The data to model ratio for the two states after fitting the \textit{NuSTAR} spectra with a disk and Comptonization component (\texttt{TBabs} * (\texttt{diskbb} + \texttt{nthComp})). The ratio clearly shows the presence of a reflection component in both the states. The Compton hump during the HSS is slightly weaker and falls off after 20 keV compared to the HIMS. Data from only one of the FPMs are shown for clarity.}
    \label{fig:ratio}
\end{figure}

We carried out a comprehensive spectral analysis of the \textit{NuSTAR}  spectra to study the evolution of the source across the two states and also constrain some physical parameters. As is typical of black hole binaries, the models constituted a thermal disk \texttt{diskbb} \citep{mitsuda84}, a thermal Comptonization \texttt{nthComp} \citep{zdziarski96, zycki99}, and a reflection component from the \texttt{relxill} family \citep{dauser14, garcia14,garcia22,dauser22}. The presence of reflection features, i.e., a broad Fe K$\alpha$ line and a Compton hump, can be seen after fitting the spectra with only the disk and Comptonization components (see figure \ref{fig:ratio}). The ISM absorption was modeled with \texttt{TBabs} where the abundance and cross sections were taken from \citet{wilms00} and \citet{verner96} respectively. The equivalent H column density ($N_H$) was fixed to $6.4 \times 10^{22}$ cm$^{-2}$ \citep{rout21a}. A constant was multiplied to the model to account for calibration differences between the two focal plane modules (FPM) of \textit{NuSTAR}. The difference between the two FPMs was found to be $\lesssim 5\%$. An absorption line around 7 keV was significantly detected in all the spectra and was fitted by a negative \texttt{gaussian} component. These represent blue-shifted absorption lines from highly ionized Fe ions present in equatorial disk wind. The key difference between the various models we experimented with is the flavour of the reflection component. 

The \texttt{xspec} notation for the total model is \texttt{constant} * \texttt{TBabs} * (\texttt{diskbb} + \texttt{nthComp} + \texttt{relxill}(X) + \texttt{gaussian}). The suffix ``X" in \texttt{relxill} represents the particular flavor opted. Henceforth, we will identify the models with the flavor of \texttt{relxill} used. While fitting, all the parameters that are not expected to change across the two spectra (from each epoch) were tied to each other and those expected to change were left free. For instance, the inner-disk temperature, photon index and the norms of all the three components were left free to vary while rest all parameters were tied across the observations \citep{xu20}.

\begin{figure}
\centering
    \includegraphics[scale=0.6]{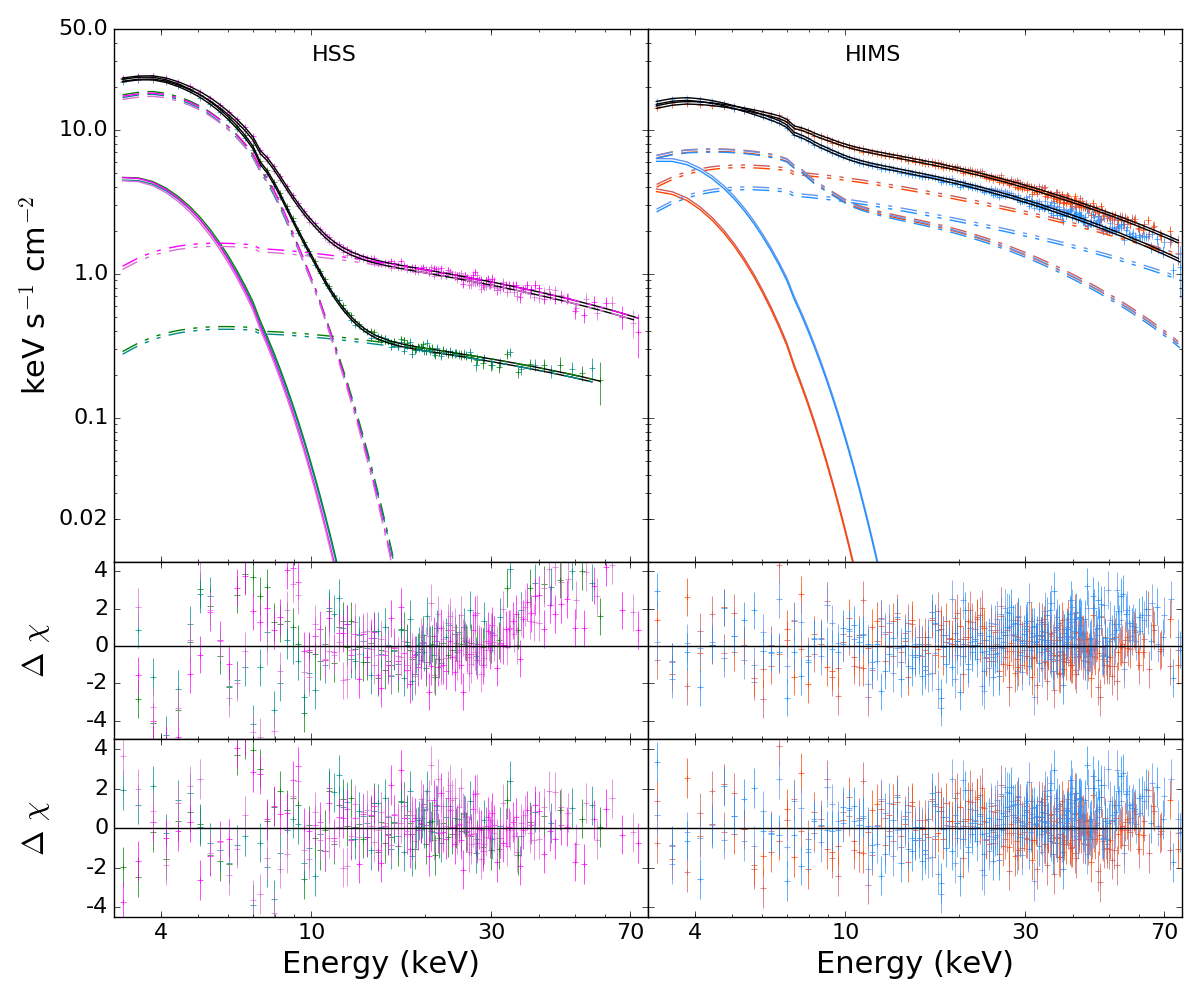}
    \caption{The spectra, model, and residuals during the HSS (green and magenta shades are for the two parts of Jan 17) and HIMS (red and blue shades are for Jan 27 and 30 respectively). The model for HSS is \texttt{constant} * \texttt{TBabs} * (\texttt{diskbb} + \texttt{nthComp} + \texttt{relxillNS} + \texttt{gaussian}) and for the HIMS is \texttt{constant} * \texttt{TBabs} * (\texttt{diskbb} + \texttt{nthComp} + \texttt{relxillCp} + \texttt{gaussian}).  Different line styles represent different model components as follows: solid - \texttt{diskbb}, dashdot - \texttt{relxill}(X), dashdotdot - \texttt{nthComp}. The black solid line represents the total model. The bottom panels show the residuals from fits with the above models. The middle panel shows the fits with a different reflection component, i.e., \texttt{relxillCp} for the HSS and \texttt{relxilllpCp} for the HIMS. The data are rebinned by a factor of 8 for clarity.}
    \label{fig:nu_specres}
\end{figure}

\subsection{HSS spectral analysis} \label{hss_ana}

We first attempted to fit the HSS spectra with both \texttt{relxilllpCp} and \texttt{relxillCp} models. The major difference between the two is that of the emissivity profiles adopted. While the former assumes a lamppost geometry characterized by the source height along the spin axis, the later considers an empirical broken power law-type emissivity. Fits to these models were quite poor with a $\chi^{2}_{\nu}$ of 1.4. There were strong residual features in the 3 - 8 keV Fe line region and an excess at energies greater than 40 keV (Figure \ref{fig:nu_specres}). In the recent release (v2.0), \texttt{relxilllpCp} has a provision for including returning radiation where a fraction of the reflected spectrum returns to the disk and results in a secondary reflection \citep{dauser22}. Including this component in our fits also did not result in any significant improvement. The one thing common in both these model flavors is that the reflection occurs when a fraction of the inverse Compton component impinges on the disk. Perhaps this power law-type irradiation is not appropriate for the spectra. 

During the HSS, the spectra of J1631 are dominated by the disk blackbody emission up to about 10 keV \citep{rout21a}. So, it could be a fraction of the high energy part of this thermal emission that results in the reflection component. \citet{cunningham76} had shown that for rapidly rotating black holes ($a > 0.9$) the thermal disk emission from the innermost stable circular orbit (ISCO) can undergo gravitational lensing and return to the disk, thus, causing reflection. This is also a form of returning radiation with a different shape which, though, is not included in the \texttt{relxill} models.  What is included, however, is a variety named \texttt{relxillNS} \citep{garcia22} where the irradiating continuum is a single temperature blackbody meant for neutron star sources. Recently, the efficiency of \texttt{relxillNS} as a proxy for returning radiation for black hole binaries has been successfully demonstrated for four sources, viz. XTE 1550--564 \citep{connors20},  4U 1630--47 \citep{connors21}, MAXI J0637--430 \citep{lazar21}, and  EXO 1846--031 \citep{wangyanan21}. Therefore, we applied the model to J1631 and it fitted quite well with a $\chi^{2}_{\nu}$ = 1.15. The best fit parameters are noted in table \ref{tab:nutable} and the spectra and residuals are shown in figure \ref{fig:nu_specres}. The spin and inclination were constrained to $> 0.996$ and $70.8^\circ$ respectively. The inner emissivity profile (q1) was found to be 9.6 and 8.5 for the two parts. Such steep emissivity profiles are not expected as true returning radiation will irradiate the outer parts of the disk and flatten the emissivity profile \citep{dauser22}. Such high values of q1 could probably arise due to the use of \texttt{relxillNS}. The accretion disk appeared to be highly dense ($\log N > 18.7$) and less ionized ($\log \xi = 1.9$) with a near solar Fe abundance ($A_{Fe}$ = 1.9). Since the flux and hardness increased in the second part of the observation, the ionization is also expected to increase. However, upon untying $\log \xi$ across the two parts also it remained consistent within errors. This indicates that the change in flux and hardness is not associated with a corresponding change in local luminosity of the disk. The absorption dip near 7 keV was fitted by a negative \texttt{gaussian}. The best fit line energy was 7.28 keV which corresponds to a wind velocity of $\sim 0.05c$ assuming the absorption to be due to Fe XXVI ions. The inner disk temperature ($Tin$) is 0.8 keV which is a bit lower than the values obtained from fits to limited bandwidth spectra that excluded a reflection component \citep{rout21a}. This is because an extra blackbody emission component ($kTbb$) is accounted within \texttt{relxillNS} \citep{lazar21} which was tied to $Tin$ in our fits. Upon keeping them untied, the difference between $Tin$ and $kTbb$ further increased with the former decreasing by about 15$\%$ and 30$\%$ in the two epochs and the later increasing marginally. The differences in the temperatures of these two thermal components, however, do not affect the best fit parameters from the reflection component, especially the spin and inclination. 

It is worthwhile to emphasize here that \texttt{relxillNS} is only used as a proxy for returning thermal radiation. It does not include the general relativistic effects that the thermal photons would actually be subjected to while returning to the disc along the geodesic. On comparison with \citet{xu20} we find that there are some differences in the results, their fits being statistically worse with $\chi^{2}_{\nu}$ of 1.3. \citet{xu20} modeled the reflection component with \texttt{refbhb} that includes interaction of disk photons in the disk atmosphere. Besides that, they include an \texttt{xstar} component for wind absorption which gives similar values for wind velocity that we obtained by fitting a negative \texttt{gaussian}. While the best fit spin for them is high ($a >0.94$) and consistent with ours, the inclination is lower ($i = 29^\circ$) compared to our fits. It is difficult to ascertain a reason for this discrepancy. Since the \texttt{relxillNS} parameters are less reliable it is even unclear if the difference in the fitted values of the inclination is real or not. This needs to be independently verified. If this difference is real one possible reason could be the different treatment of column density in the two works. \citet{xu20} leave $N_H$ to vary freely and it constrained to a low value of $3.3 \times 10^{22}$ cm$^{-2}$. The energy range of \textit{NuSTAR} is not best suited to constrain the column density which peaks at much lower energies. Therefore, we fixed $N_H$ to $6.4 \times 10^{22}$ cm$^{-2}$ obtained from \textit{NICER} fits which has a better low energy coverage \citep{rout21a}. In section \ref{hims_ana} we demonstrate how a low $N_H$ could possibly hide an absorption dip and result in a low inclination.     


\begin{table}
\centering
\caption{Best fitting parameters for fits with the HSS and HIMS spectra. The model for the HSS spectra is \texttt{TBabs}*(\texttt{diskbb}+\texttt{nthComp}+\texttt{relxillNS}+\texttt{gaussian}). For the HIMS spectra \texttt{relxillNS} was replaced by \texttt{relxillCp}. The errors represent $1 \sigma$ confidence range. All symbols have the usual meaning. The parameters with * were fixed while fitting.}

\label{tab:nutable}
\begin{tabular}{||c|c c c|c c c||}
 \hline
& & HSS & & & HIMS & \\
Parameters & 17 January (P1) & & 17 January (P2) & 27 January & & 30 January \\
 \hline\hline
 
N$_{H}$ ($\times 10^{22}\, cm^{-2}$) & & $6.4^\star$ &  & & $6.4^\star$ &  \\ 
 \hline

T$_{in}$ (keV) & $0.827 \pm 0.003 $ &  & $0.812 \pm 0.003$ & $0.75 \pm 0.01$ & & $0.84 \pm 0.01$  \\
norm$_{\text{diskbb}}$  & $3396^{+550}_{-370} $ &  & $ 3725^{+334}_{-469}$ & $4911^{+343}_{-300}$ & & $4183^{+195}_{-180}$  \\

\hline

$\Gamma$ & $2.32 \pm 0.02$ &  & $2.38 \pm 0.01$ & $2.44 \pm 0.02 $ & & $2.45 \pm 0.02$ \\
kTe (keV)& & $1000^\star$ & & & $1000^\star$  &\\
norm$_{\text{nthComp}}$ & $0.25 \pm 0.01 $ &  & $1.02 \pm 0.02$ & $4.2 \pm 0.3$ & & $2.5 \pm 0.2$ \\
 \hline

q1 & $9.6^{+0.2}_{-0.7}$ &  & $8.5^{+0.2}_{-0.7}$ & $4.9 \pm 0.2 $ &  & $4.8 \pm 0.2$ \\
q2 & & $3.0^\star$ & & & $3.0^\star$ & \\
Rbr (R$_{\text{g}}$) & & $6.0^\star$ & & & $6.0^\star$  & \\
$a$ & & $> 0.996$ & & & $0.998^\star$  & \\
Incl (degrees) & & $70.8^{+0.7}_{-1.7} $ & & & $52.5^{+2.1}_{-1.9} $  & \\
R$_{in}$ ($r_{isco}$)& & $1.0^\star$ & & & $1.35^{+0.05}_{-0.03} $  & \\
$\log \xi$ (erg cm s$^{-1}$) & & $1.90^{+0.02}_{-0.06} $ & & & $3.28 \pm 0.06$  & \\
A$_{Fe}$ & & $1.9 \pm 0.1$ & & & $1.6 \pm 0.1$  & \\
$\log$ N & & $ > 18.89$ & & & $18.4 \pm 0.1$  & \\
norm$_{\text{relxill(X)}}$ & $1.04^{+0.06}_{-0.11}$ &  & $0.94^{+0.02}_{-0.06} $ & $0.18 \pm 0.01$ & & $0.17 \pm 0.02$\\
 \hline
  
LineE (keV) & & $7.28 \pm 0.02 $ & & & $7.31 \pm 0.03$  &\\
Sigma (keV) & & $0.12 \pm 0.03$ & & & $0.29 \pm 0.03$  & \\
norm$_{\text{gauss}}$ ($\times 10^{-3}$)  & & $-1.8 \pm 0.2 $ & & & $-5.7^{+0.6}_{-0.9} $  & \\

 \hline
  
  
 \hline
$\chi^2$ (dof)  & &  3317.79 (2885) & & &  5195.49 (4927)  &\\
$\chi^{2}_{\nu}$  & & 1.15 & & & 1.05  &\\
  \hline

\end{tabular}
\end{table}


\subsection{HIMS spectral analysis} \label{hims_ana}

\begin{figure}
\centering
    \includegraphics[scale=0.6]{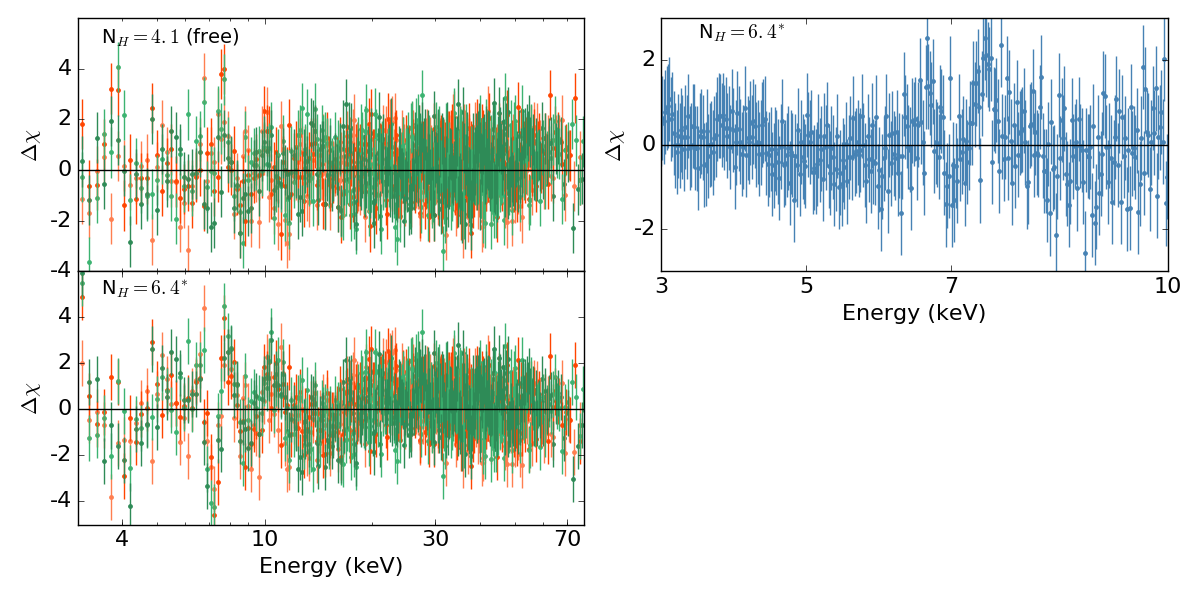}
    \caption{The top left panel shows the residuals for fits to the HIMS spectra using the exact same model opted by \citet{xu20}. The bottom left panel shows the residuals as above but with $N_H$ fixed to 6.4 $\times 10^{22}$ cm$^{-2}$. These residuals are binned by a factor of 6 for plotting. The right panel shows the residuals for a contemporary \textit{NICER} spectrum fitted to the same model. Both \textit{NuSTAR} and \textit{NICER} fits confirm to the presence of an absorption line near 7 keV which becomes apparent when the column density is fixed to 6.4 $\times 10^{22}$ cm$^{-2}$ as found by \citet{rout21a}}
    \label{fig:hims_residual}
\end{figure}

As the source transitioned to the HIMS, the spectrum became dominated by power law emission due to inverse Comptonization. Unlike the HSS, the reflection component in the HIMS would arise due to irradiation by a hard power law spectrum. Therefore we used the conventional reflection models with power law-type irradiation to study this state. To that effect we used the \texttt{relxilllpCp} and \texttt{relxillCp} flavors from the \texttt{relxill} family. Both the models gave almost equally good fits and the parameters were also consistent. For \texttt{relxilllpCp}, however, the lamppost height pegged to the minimum value of 2 R$_g$ and could not be constrained. As mentioned in section \ref{hss_ana}, very steep emissivity profiles, originating as a consequence of such low lamp height ($< 3$ R$_g$), are probably nonphysical. We were not able to find another acceptable solution with higher lamp height. Moreover, the inner emissivity profile from the \texttt{relxillCp} fits were reasonably shallow. Therefore, here we present only the results from \texttt{relxillCp} fits. The best fit parameters from fits with \texttt{relxillCp} are noted in table \ref{tab:nutable}. Figure \ref{fig:nu_specres} (right panel) shows the spectra, model, and residuals of these fits. During this state, the disk was very close to ISCO with R$_{in} = 1.35 \times $R$_{isco}$. The disk density was high ($\log N = 18.4$) and the Fe abundance was again consistent with the solar values ($A_{Fe}$=1.6) as in the HSS. The inner emissivity index was lowered to 4.9. The absorption line due to wind was constrained to 7.32 keV which again translates into out-flowing equatorial wind at 0.05$c$ assuming absorption by Fe XXVI ions. Here also, the inclination was constrained to $53.1^\circ$ which is higher than the value reported by \citet{xu20}. 

Fixing the H column density to a higher value, as discussed in section \ref{hss_ana}, had a subtle yet important implication on the fits. A possible readjustment of the continuum in the soft band results in a clear absorption line near $\sim 7$ keV which was absent in the fits by \citet{xu20}. Figure \ref{fig:hims_residual} shows a comparison of the residuals with $N_H$ left free versus it being fixed to $6.4 \times 10^{22}$ cm$^{-2}$ using the exact same model as \citet{xu20}. The bottom left panel with the higher $N_H$ clearly shows an absorption line. In order to independently verify the presence of the absorption line we analysed one \textit{NICER} observation made on 28 Jan 2019 (OBSID - 1200500114), i.e., between the two \textit{NuSTAR} observations in the HIMS. It was also the observation with the longest exposure of 10 ks and likely to show all possible features in the spectrum. The right panel in figure \ref{fig:hims_residual} shows the residuals from fitting the aforementioned \textit{NICER} observation with the model - \texttt{TBabs} * (\texttt{diskbb} + \texttt{nthComp} + \texttt{relxillCp}). The absorption feature is again clearly discernible suggesting that the line in the \textit{NuSTAR} spectrum is also real. Upon adding a negative \texttt{gaussian} component to the above model a good fit was obtained. The best fit inclination was $50.2^\circ \pm 1.8^\circ$ (90$\%$ confidence) consistent with the \textit{NuSTAR} estimates. The \textit{NICER} fits, however, have their limitation pertaining to the small energy range (0.5-10 keV) which makes constraining the power law and reflection hump difficult. 

\subsection{Black hole mass estimation} \label{mass_ana}

\begin{figure} 
\centering
    \includegraphics[scale=0.6]{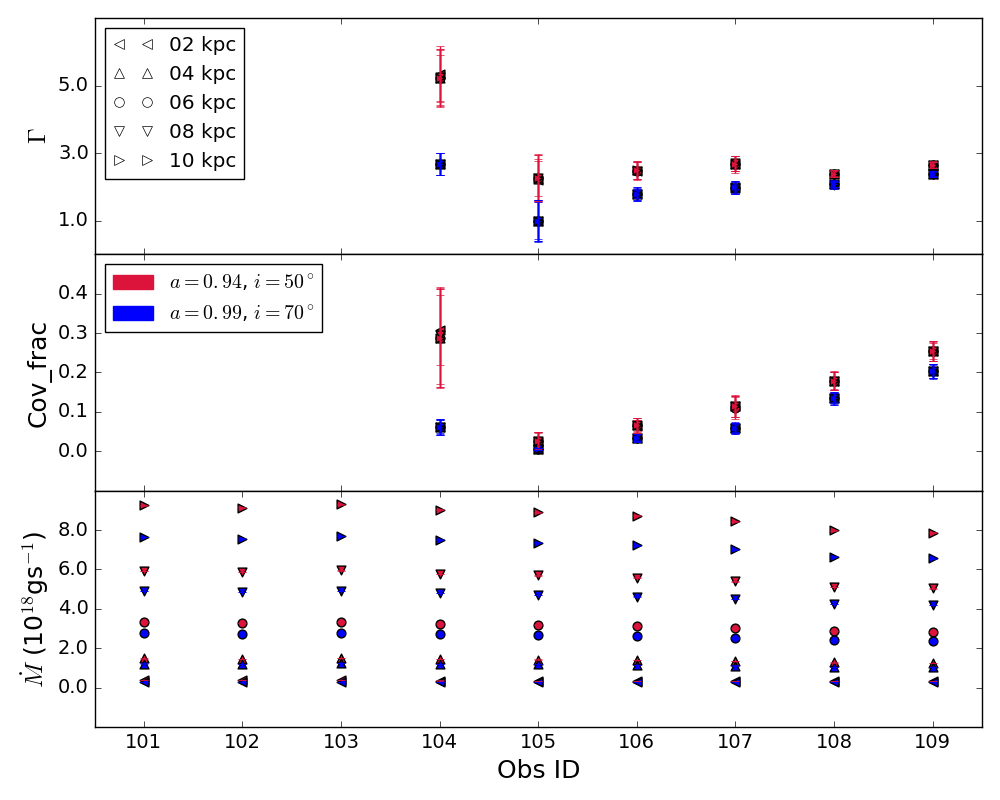}
    \caption{From top to bottom: The best-fit values of photon index, covering fraction of the corona, and accretion rate for all nine HSS spectra observed with \textit{NICER}. The labels of X-axis ticks show the last three digits of the observation ID. The different symbols corresponds to the different values of distance. The red and blue colors represent the low spin - low inclination and high spin - high inclination combination respectively. The power law component was not required for the first three observations, hence no points.}
    \label{fig:ni9bfp}
\end{figure}

A robust way of measuring the black hole mass involves calculating the binary mass function using the radial velocity of the donor star from the Doppler shifts of its absorption lines. Such a dynamic method is often inhibited by the apparent faintness of the donor star in low-mass X-ray binaries. An alternative method comprises measuring the direct X-ray flux from the accretion disk when its inner edge is at the ISCO. Fitting the soft X-ray continuum during the HSS phase of the outburst provides the ingredients for the same. As described in Section \ref{hss_ana}, the \texttt{diskbb} component in our model accounts for the thermal disk emission. Its norm is given by 
\begin{equation}
    N = (r_{in}/d_{10})^2 \cos \theta
\end{equation}
where $r_{in}$ is the apparent radius of the inner disk, $\theta$ is the inclination and $d_{10}$ is the distance to the source in units of 10 kpc. The actual radius $R_{in}$ is related to the apparent radius as $R_{in} = \epsilon \kappa^2 r_{in}$, where $\epsilon$ ($= 1.4$) is a correction factor for the inner boundary condition \citep{kubota98} and $\kappa$ ($=1.7$) is the spectral hardening factor \citep{shimura95}. From our fits to the HSS spectra the mean \texttt{diskbb} norm is found to be $3560 \pm 430$ (average 1-$\sigma$ error from both parts; see Table \ref{tab:nutable}) and the inclination is $70^\circ \pm 1^\circ$. This gives $R_{in} = (121 \pm 37)d_{10}$ km. For a kerr black hole ($a=0.998$), the ISCO is given by $1.24 R_g$ where $R_g$ is the gravitational radius. The mass of the black hole ($M$), thus, tantalizingly comes out to be 
\begin{equation}
     M = (66 \pm 20) d_{10} \: M_\odot.
\end{equation}
Even for a distance of 4 kpc, the mass will be 26 $M_\odot$ which is higher than any other stellar mass black hole in our galaxy. It must be noted here that the obtained mass is highly sensitive to uncertainties in the spin, inclination, and distance values. From our fits to the \textit{NuSTAR} spectra, both the spin and inclination are constrained to high values. A conservative estimate on mass, obtained by fixing the inclination to $50^\circ$ (from HIMS fits) and spin to 0.94 \citep{xu20}, results in $(30 \pm 15) d_{10} \: M_\odot$. 

Considering the significance of this result, we sought to explore further by analyzing the HSS data from \textit{NICER} using a physically motivated disk model and also attempt to constrain the distance using optical observations. J1631 remained in the HSS till 23 January 2019, i.e., up to \textit{NICER} observation 1200500109 from the beginning (1200500101) \citep{rout21a}. Therefore, the first nine \textit{NICER} observations were simultaneously fitted to the model \texttt{TBabs}*(\texttt{thComp}*\texttt{kerrbb}+\texttt{gaussian}) to constrain the mass. \texttt{kerrbb} is a general relativistic accretion disk model which parameterizes all the system parameters such as spin, inclination, distance, and accretion rate \citep{li05}. The convolution model \texttt{thComp} \citep{niedzwiecki19,zdziarski20a} is an improvement over \texttt{nthComp} providing higher accuracy in the mildly relativistic regime and higher optical depth, $\tau \geq 1.6$. An additional advantage is that it comptonizes the seed photons from \texttt{kerrbb}, thus, not requiring to deal with the $kT\_bb$ parameter in \texttt{nthComp}. The \texttt{gaussian} was added to account for the Fe K$\alpha$ line. While fitting, all the system parameters that do not change across the observations like column density, spin, inclination, mass, distance, and the three \texttt{gaussian} parameters were tied together. This left the covering fraction, optical depth, and mass accretion rate as untied parameters across the nine observations. The shape of the blurred Fe line is indeed not symmetric. However, considering the objective of this exercise which to characterize the intrinsic disk emission, a \texttt{gaussian} serves as a reasonable approximation for the line. The uncertainties introduced because of its use is only at a few percent level. Out of the nine, the first three observations were fully dominated by the disk flux and did not require the Comptonization component. For these three, the covering fraction were fixed to 0. The spin and inclination were fixed to a combination of extreme values. The best fit spin and inclination from the HSS fits, i.e., 0.99 and $70^\circ$, constituted the upper limits. The lower limit for spin was considered to be 0.94 \citep{xu20} and that for inclination was $50^\circ$, as obtained from our fits to the HIMS spectra. The distance was fixed to a set of values between 2 - 8 kpc. The flags for including the effects of limb darkening and disk self-irradiation were turned on. It was assumed that there would be zero torque in the inner boundary of the accretion disk. The spectral hardening factor ($\kappa$) was fixed to 1.7 \citep{shimura95}. The uncertainty in mass brought by a different choice of $\kappa$ is much smaller compared to the uncertainties due to spin, inclination, and distance. Besides, $\kappa$ is positively correlated to mass making our choice a conservative one. For each combination of spin, inclination, and distance the accretion rate, mass, covering fraction, and photon index were fitted. 

\begin{figure} 
\centering
    \includegraphics[scale=0.8]{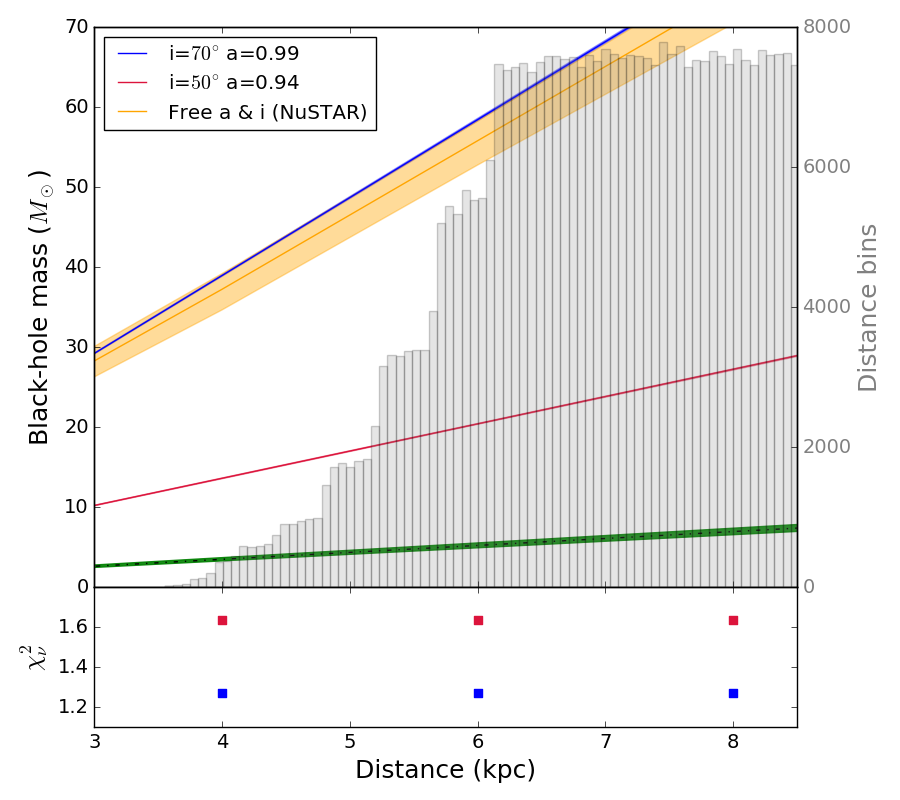}
    \caption{Top panel: The red and blue lines represent the black hole mass - distance relation obtained by simultaneously fitting nine soft state \textit{NICER} spectra for low spin - low inclination and high spin - high inclination combinations respectively. The orange line represents the same relation for fits with \textit{NuSTAR} spectra. The gray histogram represents the likelihood of distance to the source obtained by comparing the observed optical flux to that expected from an irradiated disk. The green line at the bottom is for another black hole binary 4U 1630--47 with $a = 0.90$ and $i = 65^\circ$, obtained by fitting four soft state XRT spectra. The shaded region along the lines represent 1-$\sigma$ statistical uncertainty on the mass for all except the \textit{NuSTAR} fits where the error is 90\%. Bottom panel: The $\chi^{2}_{\nu}$ values from the simultaneous \textit{NICER} fits for the two combinations of spin and inclination and for different distances.}
    \label{fig:mvsd}
\end{figure}

The results of this exercise are shown separately in figures \ref{fig:ni9bfp} and \ref{fig:mvsd}. Figure \ref{fig:ni9bfp} shows the best fit photon index ($\Gamma$), covering fraction ($cov\_frac$) and accretion rate ($\dot{M}$) for all nine observations and for the five distances. The red points represents the low spin - low inclination pair ($a=0.94$, $i=50^\circ$) and the blue points represents the high spin - high inclination pair ($a=0.99$, $i=70^\circ$). Both the \texttt{thComp} parameters ($\Gamma$ and $cov\_frac$) remain unchanged for different values of distance. It is only the \texttt{kerrbb} parameter ($\dot{M}$) which adjusts significantly; increasing from $\sim 0.1 \times 10^{18}$ g s$^{-1}$ for 2 kpc to $\sim 8.0 \times 10^{18}$ g s$^{-1}$ for 8 kpc. While \texttt{thComp} was required for the fourth observation it could not be constrained well which is apparent from the large error. 

Figure \ref{fig:mvsd} shows the variation of the best fit mass and the fit statistic ($\chi^{2}_{\nu}$) with different values of distance. The color scheme is the same as used in figure \ref{fig:ni9bfp}. The region between the red and blue lines represents the possible range of black hole mass. For other combinations, such as, high spin - low inclination and low spin - high inclination the values for mass will fall inside this region. It is quite clear that the mass range inferred here is very much consistent with the values previously calculated using the \texttt{diskbb} norm from the \textit{NuSTAR} fits. In fact, the mass - distance relation from the \textit{NuSTAR} fits alone coincides perfectly with the \textit{NICER} values (orange line). These numbers were generated by replacing \texttt{diskbb} with \texttt{kerrbb} to the \textit{NuSTAR} fits and tying the common parameters. Of additional interest is the fit statistic shown in the bottom panel. The $\chi^{2}_{\nu}$ for the low spin - low inclination combination is much worse ($\sim 1.6$) compared to the high spin - high inclination pair ($\sim 1.2$). It is also worth noting that the \texttt{thComp} parameters were better constrained for the later combination (comparatively small error, see figure \ref{fig:ni9bfp}). Moreover, the $cov\_frac$ for the fourth observation is unrealistically high for a source in the HSS with the choice of low spin - low inclination. All these suggests that the \textit{NICER} spectra prefer a high spin - high inclination combination and it consequently translates to a higher mass range.

In order to validate this method and the obtained result of high black hole mass we analyzed another source 4U 1630--47 which has close resemblance to J1631 in many aspects. It is a rapidly spinning high inclination binary showing clear presence of disk winds \citep{tomsick98,pahari18} and located in a direction very close to J1631. \citet{seifina14} have deduced the mass to be $\sim 10 \, M_\odot$ based on the correlation between photon index, low-frequency quasi-period oscillations and mass accretion rate. During the 2016 outburst, 4U 1630--47 remained predominantly in the disk dominated high soft state for most part, with the hardness only increasing slightly towards the end. The source was monitored with \textit{Swift} starting from 30 August till 21 October and then resumed on January 2017. Similar to J1631, we analysed the first four \textit{Swift}/XRT \citep{burrows05} spectra observed on 30 August and 5, 6, and 7 September during which it was in the HSS. The analysis was done using the standard procedure with adequate emphasis on accurate pile-up correction \citep{romano06,motta17}. The same procedure, as opted for J1631, was followed to find the relation between black hole mass and distance. The result of this exercise is overlaid on Figure \ref{fig:mvsd} (green line). Despite similar values of spin and inclination for both the sources, the mass ranges are clearly separated. 4U 1630--47 occupies the typical range of 3 - 10 $M_\odot$ for the entire distance range. Even if we consider the full extent of inclination, the mass range would be  significantly different in the two cases. The estimated mass of 10 $M_\odot$ \citep{seifina14} for 4U 1630--47 concurs roughly with the galactic-center distance of 8 - 10 kpc. At that typical distance, J1631 will have a mass of about $40\, M_\odot$. 

A reasonable limit on the distance could now give a possible range for the black hole mass. In order to have an estimate of the distance, we take help from the optical observations of the source. If the optical flux is assumed to be originating from the accretion disk, then by comparing the observed flux with the expected (i.e., theoretically calculated) flux some estimate of the distance can be obtained. Several attempts were made to detect the source during the outburst \citep{kong19,shin19}. However, as reported by \citet{kong19} a certain level of ambiguity exists in detecting the source at the exact location. For a star closest to the source position \citet{kong19} reported a magnitude of $19.36 \pm 0.06$ in the SDSS-r band on 10 February 2019. 18 days before this, (\citet{shin19} reported a slightly brighter object with 19.1 magnitude at the source location. This object may or may not be the optical counterpart of J1631. But the optical counterpart cannot be brighter than that, otherwise it would have been certainly detected. Therefore, the observed magnitude was considered to represent an upper limit on flux. This would give a lower limit on distance. 

We carried out a Monte Carlo simulation to infer limits on the distance. The optical flux originating from a black hole binary during outburst is a combination of emission from the irradiated outer accretion disk and the secondary star \citep[see for example][]{rout21b}. The disk flux during a typical outburst is about 2 to 5 orders of magnitude brighter than the secondary companion rendering its contribution to the total flux negligible. Therefore, the disk flux was calculated by approximating a multi-temperature black body in the geometry of an irradiated thin accretion disk \citep{frank02}. It is given by the Planck's law:
\begin{equation}
    F_\nu = \frac{4 \pi h \cos \theta \nu^3}{c^2 D^2} \int^{R_{out}}_{R_{in}} \frac{R dR}{e^{h \nu/ k T(R)} -1}
\end{equation}
where $R_{in}$ and $R_{out}$ are the inner and outer radii of the accretion disk. For a disk where the luminosity is dominated by irradiation, the temperature profile $T(R)$ becomes
\begin{equation}
    \left(\frac{T(R)}{T_{in}}\right)^4 = \frac{2}{3} \frac{H}{R_{in}}\left[\frac{d \ln H}{d \ln R} - 1\right] (1-\beta).
\end{equation}
In our simulations $\beta$, the disk albedo, was assumed to be 0.5. The scale height $H$ for a disk irradiated by X-rays from the inner accretion region is calculated from the \citet{shakura73} solution as
\begin{equation}
    H = 1.7 \times 10^8 \alpha^{-1/10} \dot{M}^{3/20} M^{-3/8} R^{9/7} \left[1 - \sqrt{\frac{R_{in}}{R}}\right]^{1/4}
\end{equation}
where $\alpha$ is the viscosity parameter (assumed to be 0.1). 

Using the above formulation the flux emitted from an accretion disk in the optical waveband (SDSS-r filter) was calculated for $10^6$ different combinations of parameters. $T_{in}$ ($0.91 \pm 0.09$ keV) was sampled from a gaussian distribution centred around the best fit value obtained by fitting the \textit{NICER} spectrum observed on the same day as the optical measurement by \citet{kong19}. Other parameters like $M$ (3 - 60 $M_\odot$), $R_{out}$ ($6\times10^4$ - $10^5 \, R_g$), $R_{in}$ (1.24 - 2.5 $R_{g}$), $D$ (1.5 - 10 kpc), and $\theta$ ($50^\circ$ - $70^\circ$) were drawn randomly from a wide uniform distribution. An essential element of this calculation, which basically constrains the distance, is extinction by interstellar dust. To incorporate this, the 3D extinction map by \citet{marshall06} was used. Since this map is in infrared band ($K_s$-filter), it was converted into SDSS-r band extinction using the conversion factors given by \citet{mathis90} and the extinction law calculated by \citet{cardelli89}. Finally, the expected magnitude (i.e., flux) was compared to the observed magnitude by \citet{kong19} for all $10^6$ iterations. Whenever the expected magnitude was more than the observed magnitude, the corresponding distance was selected. At the end of this simulation, $\approx 5.4 \times 10^5$ distances were picked. The histogram of these selected distances is plotted in figure \ref{fig:mvsd} as gray bars. The 99\% lower limit on distance is 4.5 kpc corresponding to a mass of 15 $M_\odot$. The value of mass further corresponds to the low spin - low inclination combination, which as we have shown gave a worse fit compared to the high spin - high inclination combination. On the higher side, the mass can be as high as 45 $M_\odot$. It is important to note that this limit is dependent on the total range of distance considered during simulations (which was 1.5 - 10 kpc) and the total number of iterations (which was $10^6$). If the simulation was done up to a higher distance (say 20 kpc) and a larger number of iterations, the lower limit could be higher and vice versa. Despite the unavoidable ambiguity in finding a limit, it is apparent that the mass of the black hole remains in a range higher than most other dynamically discovered masses \citep{jonker21}. Black hole X-ray binaries are mostly populated around the central parsec of the galactic centre which is about 8 kpc away \citep{generozov18}. If this was true for J1631, its mass would lie within a larger range of 30 - 70 $M_\odot$. In fact, the high H column density for the source supports this possibility of a very high distance. From the relation given by \citet{guver09}, an $N_H$ of $6.4 \times 10^{22}$ cm$^{-2}$ gives an optical extinction ($A_V$) of $\approx 29$ magnitudes \citep[or $A_R \approx 22$ magnitude;][]{cardelli89}. Such a high extinction is quite possible considering the presence of a giant molecular cloud at 11 kpc \citep{augusteijn01, tomsick14}. From the extinction map of \citet{marshall06}, one arrives at possible distances of more than 15 kpc suggesting a much higher range for mass.

\section{Discussion} \label{disc}

In this work, we re-analyzed the \textit{NuSTAR} and \textit{NICER} spectra of the galactic X-ray binary J1631 and provide newer insights on the source. The HSS spectrum was dominated by thermal emission from the disk and the reflected component originated from disk self-irradiation through returning radiation. \citet{cunningham76} demonstrated that returning radiation becomes significant for black holes with spin greater than 0.9 and inner disk reaching the ISCO. Both these criteria were satisfied by J1631 during the HSS. Furthermore, the weak direct Coronal emission ensured that reflection by irradiation of a power law spectrum is insignificant. Thus, we made use of \texttt{relxillNS}, a model originally designed for neutron stars, where the irradiating spectrum is a blackbody and it fitted the spectrum quite well. It is worth noting that \texttt{relxillNS} is only a proxy for returning radiation and does not include the complex physics associated with energy shifts of lensed photons. 

During the HIMS, on the other hand, reflection takes place by irradiation of inverse comptonized spectrum. Both \texttt{relxilllpCp} and \texttt{relxillCp} provided good fits with similar best fit parameters. The only exception being the lamppost height in the former model which pegged at 2 and could not be constrained. The best fit parameters with \texttt{relxillNS} and \texttt{relxillCp} for the HSS and HIMS spectra respectively are noted in Table \ref{tab:nutable}. Both the states were marked by the presence of ultra fast winds with  velocities approaching relativistic values (0.05$c$). The disk density was high with $\log N$ remaining  $>18.3$ and the Fe abundance was closer to solar values for both the states. The Fe abundance constraining to solar values in conjunction with a high disk density is consistent with the picture presented by \citet{tomsick18}. The ionization ($\log \xi$) during the HSS was 1.92 erg cm s$^{-1}$ and that during the HIMS was 3.28 erg cm s$^{-1}$. This is also consistent with the idea that disk ionization increases with increase in source hardness \citep{done07}. The inner emissivity index decreases from a steep value of $\sim 9$ to a 4.9 while transitioning across the states. The black hole in J1631 is found to be spinning at a near maximal value and the disk inclination is constrained to lie between $53^\circ$ - 70$^\circ$. 



While the high spin estimate is consistent with the previous report by \citet{xu20}, the high inclination is not. In Section \ref{hims_ana} we demonstrated that this discrepancy in inclination could arise from a low $N_H$ obtained by leaving it free. The disk inclination is an important parameter as it strongly affects the inferred black hole mass. Therefore, we discuss a few other arguments that independently support a high inclination for J1631. \citet{munozdarias13} studied the outburst evolution of 11 black hole binaries and concluded that the high inclination sources follow a triangular track on the hardness-intensity plot while low inclination sources follow a more squarish track. J1631 does indeed follow such a triangular track on the HID indicating a higher inclination angle \citep{monageng21}. Over the last several years there has been an increasing consensus on the geometrical origin of the Type C QPOs \citep{ingram09,ingram11}. The precessing hot flow model explains the dynamic origin of the QPOs quite well \citep{ingram15, cheng19}. \citet{you18} carried out extensive simulations under the framework of Lense-Thirring precession model to predict the fractional variability spectrum during different states of an outburst and for various physical parameters. The broadband rms spectra of low frequency quasi periodic oscillations in J1631 calculated with \textit{Insight}/HXMT data matches well with simulations for high inclination sources \citep{bu21}. The rms increases with energy below 10 keV and then flattens above 10 keV as predicted by \citet{you18} and also observed in other high inclination sources such as GRS 1915+105 \citep{rodriguez04}, XTE J1859+226 \citep{casella04}, H1743--322 \citep{li13}, etc. The phase lag between hard and soft photons at the QPO frequency is also known to show significant inclination dependence \citep{vandeneijnden17}. While low inclination sources show hard lag, high inclination sources show soft lags for QPOs with frequencies $>2$ Hz. Recently, \citet{wang22} have reported the detection of reverberation (soft) lags in 9 out of 12 good data groups from \textit{NICER} observations of J1631. This also suggests that the accretion disk in J1631 is likely to be at a higher inclination. After a comprehensive study of ionized winds in galactic black hole binaries using high resolution X-ray spectra, \citet{ponti12} demonstrated that such winds should be, and are indeed ubiquitously, found in high inclination binaries. The presence of relativistic and ionized wind signatures in J1631 is yet another piece of evidence independently supporting the thesis for a high inclination. 


Once the spin and inclination are robustly constrained, the mass of the black hole can be obtained from the disk emission during the HSS \citep{parker16}. As discussed in section \ref{mass_ana} we simultaneously fitted 9 spectra of J1631 from \textit{NICER} using the general relativistic disk model \texttt{kerrbb} for different combinations of spin and inclination. As shown in figure \ref{fig:mvsd} the mass of the black hole remains high for a wide range of distances and different combinations of spin and inclination. An expected range of distances was also calculated by simulating the emission from an irradiated accretion disk and then comparing the expected flux in the optical band with the observations. The gray histogram in figure \ref{fig:mvsd} represents the likelihood of possible distances. This histogram gives a 99\% lower limit on distance at 4.5 kpc which translates into a mass range of 15 - 45 $M_\odot$. The high $N_H$ and consequently the extinction suggest that the distance to source could be much higher than 15 kpc which will put the mass around 50 $M_\odot$. The mass is poorly constrained owing to the uncertainty in the spin and inclination estimate. The lower and upper limits on mass corresponds to the spin and inclination combinations of (0.94, $50^\circ$) and (0.99, $70^\circ$) respectively. These estimates were derived from fits to the \textit{NuSTAR} spectra using reflection spectroscopy. As ambiguity may still exist in the reflection results, especially in the inclination, we explored and have discussed independent avenues supporting a high inclination consistent with our reflection fits. It was further shown using fits to the \textit{NICER} spectra that they prefer a high spin and high inclination combination (Figure \ref{fig:mvsd}). In order to validate that the intrinsic disk flux for J1631 does indeed favour a high mass, the same exercise of fitting the HSS spectra with \texttt{kerrbb} was repeated for another well-known system 4U 1630--47. The expected values of mass for this source in the entire range of distances lied below 10 $M_\odot$. The green line in figure \ref{fig:mvsd} represents the mass-distance relation for 4U 1630--47.  

If our mass estimate is true, the black hole in J1631 is possibly the most massive in our galaxy found in X-ray binaries. Therefore, we strongly recommend optical/infrared follow-up of the secondary star during quiescence which can robustly measure the black hole mass. Claims for massive stellar-mass black holes have mostly not stood the test of time. An extra-galactic system IC 0 X--1 was thought to host a 30 $M_\odot$ black hole \citep{silverman08}. However, it was later found that the radial velocity curve of He II emission line originated from stellar wind of the Wolf-Rayet companion and did not trace the binary's orbital motion \citep{laylock15,binder15}. Similarly, the claim of a 70 $M_\odot$ black hole in LB-1 \citep{liu19} was also later refuted \citep{abdulmasih20,elbadry20}. Till now, the most massive galactic stellar-mass black hole in X-ray binaries is Cygnus X--1 \citep[20 $M_\odot$;][]{millerjones21}. A massive stellar-mass black hole is quite interesting for various reasons. Not only does it pose challenges for stellar and binary evolution theories, it also puts to test the apparent dichotomy in black hole masses discovered with electromagnetic and gravitational waves (GW) \citep{perna19}. For example, in a recent study \citet{jonker21} have shown that the observed black hole mass distribution in X-ray binaries is biased against massive black holes. They found that most discovered black holes are located at a larger distance from the galactic center and also at a higher height from the galactic plane. Most star forming regions, where massive black holes ($>20\, M_\odot$) are expected to form, remain close to the galactic plane. Sources along these directions suffer heavy dust extinction, thus, inhibiting their dynamic mass measurement. Incidentally, J1631 is situated close to the galactic center and along the galactic plane ($l=336.3^\circ, \, b=0.3^\circ$) which is commensurate with the high column density derived in \citet{rout21a}. This makes J1631 an ideal, and likely the first, candidate for a massive stellar-mass black hole in an X-ray binary with a mass in the range of its GW counterparts.

\section{Summary}

A comprehensive spectroscopic analysis of the galactic X-ray binary J1631 was carried out using data from \textit{NuSTAR} and \textit{NICER} observatories. The main results of this work are summarised below:

\begin{itemize}
    \item The HSS spectrum was dominated by thermal emission from the accretion disk. This strong thermal emission also acted as the source for reflection by gravitationally bending and self-irradiating the disk. 
    \item The transition from the HSS to HIMS was marked by an increase in the Comptonization component, which also became the dominant source for reflection unlike the HSS. The increase in power law fraction also reflected in an increase in disk ionization ($\log \xi$ increasing from 1.9 to 3.3). 
    \item The disk density remained high ($\log N \geq 18.4$) and the Fe abundance was close to solar values during both the states ($A_{Fe} \approx 1.8$). 
    \item Both the spin of the black hole ($a > 0.996$) and inner disk inclination ($50^\circ - 70^\circ$) were constrained to high values.
    \item The mass of the black hole was found to be very high, with a 99\% lower limit of 15 $M_\odot$ for a distance of 4.5 kpc.
\end{itemize}

\begin{acknowledgments}
This research is supported by Physical Research Laboratory which is funded by Dept. of Space, Govt. of India. This research has made use of data and software provided by the High Energy Astrophysics Science Archive Research Center (HEASARC), which is a service of the Astrophysics Science Division at NASA/GSFC.
\end{acknowledgments}


\bibliography{references}{}
\bibliographystyle{aasjournal}



\end{document}